\documentclass[10pt]{iopart}

\usepackage{graphicx,iopams,braket,bm,color,setstack,hyperref,braket}

\newcommand{\Var}{\mathop{\mathrm{Var}}\nolimits}
\newcommand{\Cov}{\mathop{\mathrm{Cov}} \nolimits}
\newcommand{\re}{\mathop{\mathrm{Re}}\nolimits}
\newcommand{\im}{\mathop{\mathrm{Im}} \nolimits}
\newcommand{\openone}{\leavevmode\hbox{\small1\normalsize\kern-.33em1}}

\begin{document}

\title{Rotation sensing at the ultimate limit}

\author{Aaron~Z~Goldberg$^{1}$, Andrei~B~Klimov$^{2}$, 
Gerd~Leuchs$^{3,4}$ and Luis~L~S\'anchez-Soto$^{3,5,*}$}

\address{$^{1}$ Department of Physics, University of Toronto, Toronto, Ontario~M5S~1A7, Canada}
\address{$^{2}$ Departamento de Fisica, Universidad de Guadalajara, 44420~Guadalajara, Jalisco, Mexico}
\address{$^{3}$ Max-Planck-Institut f\"ur die Physik des Lichts,  91058~Erlangen, Germany}
\address{$^{4}$ Institute of Applied Physics, Russian Academy of Sciences, 603950~Nizhny Novgorod, Russia}
\address{$^{5}$ Departamento de \'Optica, Facultad de F\'{\i}sica, Universidad Complutense, 28040~Madrid, Spain}

\ead{lsanchez@fis.ucm.es}
\vspace{10pt}
\begin{indented}
\item[]\today
\end{indented}

\begin{abstract}
Conventional classical sensors are approaching their maximum sensitivity levels in many areas. Yet these levels still are far from the ultimate limits dictated by quantum mechanics. Quantum sensors promise a substantial step ahead by taking advantage of the salient sensitivity of quantum states to the environment. Here, we focus on sensing rotations, a topic of broad application. By resorting to the basic tools of estimation theory, we derive states that achieve the ultimate sensitivities in estimating both the orientation of an unknown rotation axis and the angle rotated about it. The critical enhancement obtained with these optimal states should make of them an indispensable ingredient in the next generation of rotation sensors that is now blossoming.
\end{abstract}

\eqnobysec
%
\vspace{2pc}
\noindent{\it Keywords}: quantum sensing, estimation theory, rotation sensors, ultimate limits
%
%
%
%
\section{Introduction}

Precise rotation sensing is a critical requisite for applications as diverse as inertial navigation~\cite{Grewal:2013aa,Lawrence:1998aa}, geophysical studies~\cite{Stedman:1997aa}, and tests of general relativity~\cite{Leuchs:1986aa,Cerdonio:1988aa,Ciufolini:1998aa}, to cite but a few. 

Since the first gyroscope, invented in 1852 by Foucault as an extension of his famous pendulum used to demonstrate the Earth's rotation, rotation sensing has experienced a formidable transformation.  Depending on the application and its environmental conditions, various types of sensors have been reported. Some of the most commonly used are based on Sagnac interferometers~\cite{Lefere:2014aa}, although they suffer from limited sensitivity and stability. Matter-wave interferometers have demonstrated superior performance~\cite{Gustavson:2000aa,Durfee:2006aa,Savoie:2018aa}, but at the price of a substantial increase in complexity.  Trapped atoms in a guiding potential have also been proposed to improve rotation precision~\cite{Wu:2007aa,Moan:2020aa}.

Irrespective of the technique, sensors are reaching performance levels where quantum effects come into play~\cite{Degen:2017aa} and, accordingly, should be re-examined from a full quantum perspective: this is the main goal of this paper. Actually, quantum metrology ascertains the ultimate bounds on the achievable measurement precision and identifies states that would be optimal for those measurements~\cite{Giovannetti:2011aa}. The main tool for those tasks is the quantum Cram\'er-Rao bound, which bounds the covariance matrix between parameters being estimated by the inverse of the quantum Fisher information matrix~\cite{Braunstein:1994aa,Braunstein:1996aa}. 
 
When the axis of the rotation is known, the only endeavour is to estimate the angle of rotation. This is entirely equivalent to determining a phase in interferometry~\cite{Rafal:2015aa}, a paradigmatic example of single-parameter estimation. This topic is well understood~\cite{Belavkin:1976aa,Helstrom:1976ij}: classical states have a precision limited by shot noise, leading to uncertainties scaling as $1/\sqrt{N}$, where $N$ is the number of particles involved in the measurement~\cite{Braginskii:1975aa,Caves:1980aa}. However, one can find special quantum states saturating the quantum Cram\'er-Rao bound: they achieve the so-called Heisenberg limit, in which measurement uncertainties scale as $1/N$~\cite{Dowling:1998aa,Sanders:1995aa,Bollinger:1996aa,Lee:2002aa,Mitchell:2004aa,Berry:2009aa}.  

In general, however, a rotation is characterized by three parameters \cite{Grafarend:2011aa}: either the two angular coordinates of the rotation axis and the angle rotated around that axis, or the Euler angles~\cite{Diebel:2006aa}. We thus face the problem of simultaneously estimating of multiple parameters, which has been considered as a distinguished feature combining classical and quantum aspects of uncertainty: the noncommutativity of quantum theory leads to nontrivial tradeoffs that are absent in classical and in single-parameter estimation problems. This observation led to a development of new multiparameter bounds that constitute the backbone of a new and lusty research line~\cite{Paris:2009aa,Szczykulska:2016aa,Braun:2018aa,Sidhu:2020aa,Albarelli:2020aa,Polino:2020aa,Rafal:2020aa}.

In this paper, we discuss optimal states for the simultaneous estimation of the three parameters of a rotation. These states attain the Heisenberg limit, confirming that they always achieve a significant enhancement in measurement precision relative not just to the classical case but also to the best single parameter schemes. Finally, we briefly discuss their practical implementation.

\section{Classical estimation}

\begin{figure}[t]
  \centering{\includegraphics[width=.95\columnwidth]{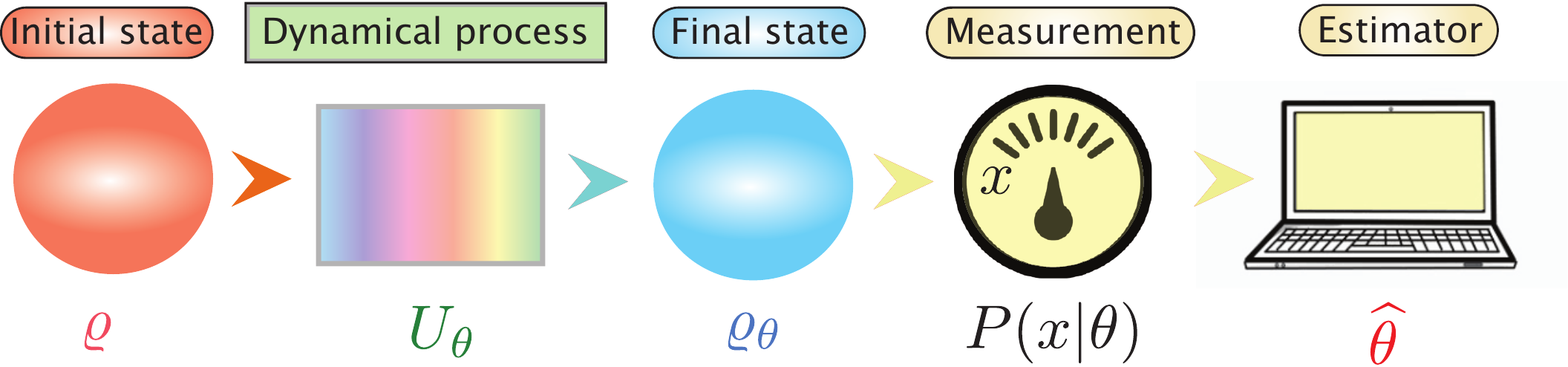}}
  \caption{Scheme of a parameter estimation protocol. A probe is prepared, then it interacts with a dynamical process that imprints the unknown parameter $\theta$ onto the probe. The resulting state, encoding the information about $\theta$, is measured and generates outcomes $x$. Based on the outcomes $x$, a suitable estimator $\widehat{\theta}$ provides an estimate of the parameter $\theta$.}
  \label{fig:fig1}
\end{figure}

Estimation theory deals with devising schemes that extract as precisely as possible the value of an unknown parameter and therefore lies in the realm of metrology. From a physical perspective, a typical estimation process can be roughly divided into three stages, which are schematized in figure~\ref{fig:fig1}: probe preparation, interaction with the system, and probe readout.  

The measured data are always affected by noise, so they are effectively represented by a stochastic variable. In some cases, the experiment cannot be modeled mathematically and the use of nonparametric estimation is necessary~\cite{Tsybakov:2009aa}. However, it is always more efficient to find a convenient model for the studied experiment. Such a parametric estimation will be the main focus of this paper. 

In classical (or frequentist) estimation, the unknown parameter is taken to be deterministic and constant during the experiment: randomness is thus solely due to noise. In contradistinction, Bayesian estimation assumes that the unknown parameter is itself a random variable distributed according to some prior probability distribution.  Whenever this distribution is unknown, it is safer to perform classical estimation, which will be our primary goal.

\subsection{Single-parameter estimation}

Let us first consider the problem of estimating the value of a single parameter $\theta$ from the observed measurement result $x$, where the latter is taken to be a continuous random variable (the discrete case can be dealt with much in the same way). In general, experiments consist of $N$ trials yielding a set of outcomes $\mathbf{x} = (x_{1},\ldots, x_{N})^{\top}$, where the superscript $\top$ denotes the transpose. The inference of the parameter is related to the measurement outcomes through some conditional probability density that is dictated by a model of the process and that we denote by $P(\mathbf{x} | \theta)$. The $N$ random variables are typically assumed to be independent and identically distributed (iid). Identically distributed implies that each of these $N$ random variables is governed by the same parameter $\theta$, whereas independence implies that the joint distribution of all these $N$ random variables is given by a product:
\begin{equation}
P(\mathbf{x} | \theta ) = \prod_{i=1}^{N} P(x_{i}|\theta )\, .
\end{equation}

To extract the parameter $\theta$ from the data $\mathbf{x}$ we use an estimator $\widehat{\theta} (\mathbf{x})$, which is a function of the observed data only. This estimator is said to be unbiased if its mean value coincides with the true value of the unknown parameter (i.e., there is no systematic error in the estimation)
\begin{equation}
\mathbb{E}_{\mathbf{x}} [ \widehat{\theta}  ] = \theta \, ,
\end{equation}
where we denote by
\begin{equation}
\mathbb{E}_{\mathbf{x}} [ \widehat{\theta} ] = 
\int \rmd \mathbf{x} \, P(\mathbf{x} | \theta) \, 
\widehat{\theta} (\mathbf{x})
\end{equation}
the expectation value of the estimator $\widehat{\theta}(\mathbf{x})$ with respect to the probability distribution $ P(\mathbf{x} | \theta)$.

A vast number of estimators have been formulated by proposing \emph{ad hoc} functions $\widehat{\theta}(x)$. The reader is referred to the excellent textbooks on parametric estimation~\cite{Trees:1968aa,Scharf:1991aa,Kay:1993aa}. One of the most widely adopted is the maximum likelihood (ML) estimator, defined as
\begin{equation}
\widehat{\theta}_{\mathrm{ML}} (\mathbf{x}) = 
\arg \max_{\theta} P(\theta | \mathbf{x}) \, .
\end{equation}
Here, the likelihood $P(\theta|\mathbf{x})$ must be interpreted  as the probability of $\theta$ being the true value given the data set $\mathbf{x}$. This estimator is unbiased in the asymptotic limit of many independent samples. Actually, it just picks the value of $\theta$  that makes $\mathbf{x}$  the most likely observation; in other words, \emph{it bets on the winner.}

A natural figure of merit to quantify the performance of an estimator is its mean square error (MSE)
\begin{equation}
\mathrm{MSE}_{\mathbf{x}} (\widehat{\theta}) = 
\mathbb{E}_{\mathbf{x}} [ (\widehat{\theta} - \theta)^{2}] \, .
\end{equation}
This definition depends on the unknown parameter $\theta$, but the MSE is \emph{a priori} a property of the estimator. Actually, it can be decomposed as
\begin{equation}
\mathrm{MSE}_{\mathbf{x}} (\widehat{\theta}) = \Var_{\mathbf{x}} (\widehat{\theta}) + 
\mathrm{Bias}^{2}_{\mathbf{x}} (\widehat{\theta}) \, , 
\end{equation}
where the variance and the bias are, respectively,
\begin{equation} 
\Var_{\mathbf{x}} (\widehat{\theta}) = \mathbb{E}_{\mathbf{x}} \left [ \left ( \widehat{\theta} -  \mathbb{E}_{\mathbf{x}} [ \widehat{\theta}]\right )^{2} \right ] \, , 
\qquad
\mathrm{Bias}^{2}_{\mathbf{x}} (\widehat{\theta}) =  \left ( 
\mathbb{E}_{\mathbf{x}} [ \widehat{\theta} ] - \theta \right )^{2}   \, . 
\end{equation}
The minimum MSE estimator finds a trade-off between the bias and the variance for every value of  $\theta$. Unfortunately, the bias is often a function of $\theta$ and, consequently, the minimum MSE estimator is generally not realizable. In general, any estimator depending on the bias will be unfeasible. This limitation prompts one to focus uniquely on unbiased estimators, with the resulting estimator usually referred to as the minimum variance unbiased (MVU) estimator. 

There are procedures for finding the MVU estimator~\cite{Kay:1993aa}. Unfortunately, they are often tedious and sometimes fail to produce the MVU estimator, whose existence is not even guaranteed. Fortunately, the ML estimator is known to approximately provide the MVU estimator under mild regularity conditions.

A fundamental tool to characterize the achievable bounds on estimation uncertainties is the Fisher Information (FI)~\cite{Fisher:1925aa}, which captures the amount of information encoded in the output probabilities. For the random variable $x$, it is defined by 
 \begin{eqnarray}
\mathsf{F}_x (\theta) & = & \mathbb{E}_{x} 
\left[ \left ( \frac{\partial \ln P(x|\theta)}
{\partial \theta} \right )^{2} \right ] = 
\int \rmd x \,  P(x | \theta) 
\left [ \frac{\partial  \ln P(x|\theta)}
{\partial \theta} \right ]^{2}
\nonumber \\
& = & 
\int \rmd {x} \,  \frac{1}{P(x|\theta)}
\left [ \frac{\partial  P(x |\theta)}
{\partial \theta}  \right ]^{2} \, ,
\end{eqnarray}
where the alternative form in the second line can be obtained by a direct computation. We keep the subscript $x$ as a reminder that the FI hinges upon the data $x$, which in fact depends on both the probe state and the measurement. The choice of the function $\ln P(x|\theta)$ is meaningful because it is additive for independent samples due to the logarithm. 

Intuitively, the FI quantifies the sensitivity of a system to a change in $\theta$: a larger amount of information is associated with larger variations of the output probabilities. This intuition was formalized with a time-honoured result called the Cram\'er-Rao bound (CRB)~\cite{Rao:1945aa,Cramer:1946aa}. It links $\mathsf{F}_{\mathbf{x}}(\theta)$ with the ultimate bound achievable by the variance of any arbitrary unbiased estimator, with $N$ identical and independent probes and measurements:
\begin{equation}
\label{eq:CRB1}
\Var_{\mathbf{x}} (\widehat{\theta}) \geq \frac{1}{N \, \mathsf{F}_{x} (\theta)} \, .
\end{equation}
The CRB applies only to well behaved probability distributions that satisfy the following regularity condition~\cite{Kay:1993aa}
\begin{equation}
\label{eq:regul}
\mathbb{E}_{\mathbf{x}} \left[  
\frac{\partial \ln P(\mathbf{x}|\theta)}{\partial \theta}  
\right ] = 0 \, .
\end{equation}
An estimator that saturates the inequality (\ref{eq:CRB1}) is said to be efficient, which can be guaranteed by the condition
\begin{equation}
\widehat{\theta} (\mathbf{x}) - \theta = \frac{\partial \ln P(\mathbf{x}|\theta)}{\partial \theta} \, .
\end{equation}
In the limit of a large number of measurements, the ML  estimator is efficient, for its distribution normally converges to the real value with a variance that saturates the CRB. 

The derivation of the CRB relies on the Cauchy-Schwarz inequality using $\partial \ln P(\mathbf{x}|\theta) / \partial \theta$, which is called the score function, as the basic object. Other choices of the score give different bounds, such as the Hammersley–Chapman–Robbins~\cite{Hammersley:1950aa,Chapman:1951aa}, Bhattacharya~\cite{Bhattacharya:1966aa}, and Barankin~\cite{Barankin:1949aa} bounds. 

To conclude, we point out that in assessing the performances of an estimator, a relevant factor is the scaling of the variance with the mean value rather than the absolute value of the variance. This feature may be quantified by means of the signal-to-noise ratio~\cite{Paris:2009aa} (for a single detection event)
\begin{equation}
R_{x} (\widehat{\theta}) = 
\frac{\widehat{\theta}^{\;2}}{\Var_{\mathbf{x}} (\widehat{\theta})} \, ,
\end{equation}
which is larger for better estimators. Using the CRB, one easily derives that the signal-to-noise ratio of any estimator is bounded by the quantity $R_{x} (\widehat{\theta}) = \widehat{\theta}^{\,2} \mathsf{F}_{x} (\theta)$. The parameter $\theta$ is effectively estimable when the corresponding $R_{x} (\widehat{\theta})$ is large.

\subsection{Multiparameter estimation}

The central task is now estimating a series of $D$ unknown parameters  $\btheta = (\theta_{1}, \dots, \theta_{D} )^{\top}$, which are assessed through a set of estimators $\widehat{\btheta} = (\widehat{\theta}_{1}, \ldots, \widehat{\theta}_{D} )^{\top}$, after the measurement $\mathbf{x}= (x_{1}, \ldots, x_{D})^{\top}$. As before, we assume $N$ trials that give $N$ distinct iid random variables $\mathbf{x}$. Each parameter $\theta_{i}$, with $i = 1,\ldots, D$, represent a physical quantity. The figure of merit is now the covariance matrix, with elements ($i,j = 1, \ldots, D$)
\begin{equation}
[ \Cov_{\mathbf{x}} ( \widehat{\btheta} )]_{ij} = \mathbb{E}_{\mathbf{x}}
\left [ 
( \widehat{\theta}_{i} - \mathbb{E}_{\mathbf{x}} [ \widehat{\theta}_{i} ] ) 
( \widehat{\theta}_{j} - \mathbb{E}_{\mathbf{x}} [ \widehat{\theta}_{j} ] )
\right ] \,   \, ,
\end{equation}
which quantifies the sensitivity relative to each parameter, while also taking into account the possible correlations between the parameters.

Much in the same way, the FI is generalized to the FI matrix $\bm{\mathsf{F}}_{\mathbf{x}} (\btheta)$ with elements
\begin{eqnarray}
[\mathsf{F}_{\mathbf{x}} ( \btheta ) ]_{ij} & = &  \int \rmd \mathbf{x} \;  
P(\mathbf{x} | \btheta) \frac{\partial \ln P(\mathbf{x}|\btheta)}
{\partial \theta_{i}} 
\frac{\partial \ln P(\mathbf{x}|\btheta)}
{\partial \theta_{j}} 
 \nonumber \\
& = &  
\int \rmd \mathbf{x} \, \frac{1}{P(\mathbf{x}|\btheta)} \; 
\frac{\partial P(\mathbf{x}|\btheta)}{\partial \theta_{i}}
\frac{\partial P(\mathbf{x}|\btheta)}{\partial \theta_{j}} \, ,
\end{eqnarray}
which is symmetric and positive. This matrix can be interpreted as a metric on the parameter space and transforms as~\cite{Sidhu:2020aa}
\begin{equation}
    \bm{\mathsf{F}}_{\mathbf{x}} ( \bm{\vartheta}) = 
    \mathbf{J}(\btheta,\bm{\vartheta})^\top \;
    \bm{\mathsf{F}}_{\mathbf{x}} (\btheta ) \; 
    \mathbf{J}(\btheta,\bm{\vartheta}) \, , 
    \label{eq:FI change of param}
\end{equation} 
for Jacobian matrix $J_{ij}(\btheta,\bm{\vartheta}) =  {\partial \theta_i}/{\partial \vartheta_j}$ governing the transformations between parametrizations.

For an unbiased estimator, the CRB in the multiparameter case is generalized to the following matrix inequality 
\begin{equation}
\bm{\Cov}_{\mathbf{x}} ( \widehat{\btheta} )  \ge \frac{1}{N} 
\bm{\mathsf{F}}^{-1}_{\mathbf{x}} ( \btheta ) \, . 
\end{equation}

An important example is the Gaussian (normal) distribution, which allows for a closed form of $\bm{\mathsf{F}}$. If the probability distribution is characterized by mean values $\bmu$ and covariance matrix $\bSigma$, 
\begin{equation}
P (\mathbf{x}|\btheta) = \frac{1}{\sqrt{(2 \pi)^{D} \det \bSigma}} 
\exp \left [ - \case{1}{2} (\mathbf{x} - \bmu)^{\top} \bSigma^{-1} 
(\mathbf{x} - \bmu) \right ] \, ,
\end{equation}
the FI matrix reads
\begin{equation}
\bm{\mathsf{F}}_{\mathbf{x}} ( \btheta) = 
\frac{\partial \bmu^{\top}}{\partial \theta_{i}} \, \bSigma^{-1} 
\frac{\partial \bmu}{\partial \theta_{j}} + 
\frac{1}{2} \Tr \left (  \bSigma^{-1} 
\frac{\partial \bSigma }{\partial \theta_{i}}
\;  \bSigma^{-1} 
\frac{\partial \bSigma }{\partial \theta_{j}}
\right ) \, .
\end{equation}

\section{Quantum estimation}

We next review the quantum setting of the problem. The general scheme is illustrated at the bottom of figure~\ref{fig:fig1}. From a quantum perspective, this protocol involves a two-step optimization problem: one must first make a smart choice of the probe state that is sensitive to the parameter, and then pick an appropriate measurement that maximizes the information extracted from the probe.

\subsection{Single-parameter estimation}

We represent the probe state by a density operator $\varrho$.  The  single parameter $\theta$ is encoded via a quantum channel $\mathcal{E}_{\theta}$, whose action on the state $\varrho$ is the transformation $\mathcal{E}_{\theta} [ \varrho ] = \varrho_{\theta}$~\cite{Chuang:2000aa}. For simplicity, we will restrict our attention to unitary channels, so that $\varrho_{\theta} = U_{\theta} \; \varrho \; U_{\theta}^{\dagger}$, with the unitary operator $U_{\theta}$ expressed as  
\begin{equation}
U_{\theta}= \exp (- \rmi \theta \, G) \, , 
\end{equation}
where $G$ is a selfadjoint operator that is called the generator of the  transformation.

In order to estimate $\theta$, we perform a measurement that is represented by some positive operator-valued measure (POVM) $\{ \Pi_{\mathbf{x}} \}$. The latter comprise a set of positive semidefinite, selfadjoint operators that resolve the identity~\cite{Holevo:2003fv}; that is,
\begin{equation}
\Pi_{\mathbf{x}} \ge 0 \, , 
\qquad
\Pi^{\dagger}_{\mathbf{x}} = \Pi_{\mathbf{x}} \, ,
\qquad
\int d\mathbf{x} \; \Pi_{\mathbf{x}} = \openone \, .
\end{equation}
By performing this measurement, we obtain a statistical distribution that, according to Born's rule, is given by
\begin{equation}
\label{eq:Born}
P(\mathbf{x} | \theta) = \Tr ( \varrho_{\theta} \, 
\Pi_{\mathbf{x}}) \, .
\end{equation}
Afterward, what remains is to obtain the best estimate of $\theta$ given  $P(\mathbf{x} | \theta)$. This can be accomplished with the basic tools outlined in the previous section, albeit one has to guarantee the positivity of the quantum state. In other words, quantum estimation can be seen just as classical estimation supplemented with the constraints imposed by positivity.  

There is an infinite number of possible POVMs that we can consider. It is therefore natural to ask whether there is an optimal measurement that should be performed on $\varrho_{\theta}$. The quantum Fisher information (QFI) is defined precisely for this purpose~\cite{Petz:2011aa}:
\begin{equation}
\mathsf{Q}_{\varrho} (\theta ) = \sup_{ \{ {\Pi}_{\mathbf{x}} \} }  
\mathsf{F}_{\mathbf{x}} (\theta) \, ,
\end{equation}
which, as indicated by its subscript, depends exclusively on the initial probe state $\varrho$. By its very same definition, we have $\mathsf{Q}_{\varrho} ( \theta ) \geq \mathsf{F}_{\mathbf{x}} (\theta)$ and so we have a quantum Cram\'er-Rao bound (QCRB) given by
\begin{equation}
\label{eq:QCRB}
\Var_{\varrho} (\widehat{\theta}) \geq \frac{1}{N \; 
\mathsf{Q}_{\varrho} (\theta )} \, .
\end{equation}
Here, the variance must be computed using the probability density (\ref{eq:Born}) associated with Born's rule. The right hand side of (\ref{eq:QCRB}) thus represents the ultimate achievable precision regardless of the measurement.

Helstrom, using fairly elementary arguments~\cite{Helstrom:1976ij}, showed an explicit way to compute $\mathsf{Q}_{\varrho} ( \theta )$; it reads 
\begin{equation}
\mathsf{Q}_{\varrho} ( \theta ) = 
\Var_{\varrho_{\theta}} ( L_{\theta} ) =  
\Tr (\varrho_{\theta} \; L^{2}_{\theta}) \,,
\end{equation}
where the variance of an operator is $\Var_{\varrho} (L) = \Tr (\varrho L^{2}) - [\Tr (\varrho L)]^{2}$, $L_{\theta}$ is the so-called symmetric logarithmic derivative, defined implicitly via
\begin{equation}
\frac{\partial {\varrho}_{\theta}}{\partial \theta} = 
\case{1}{2} \{ \varrho_{\theta}, L_{\theta} \} \, , 
\end{equation}
and $\{ \cdot, \cdot \}$ stands for the anticommutator $\{ A, B \} = A B + B A$.

The QFI  has a number of interesting properties. First, it is convex in the quantum states: given any two states $\varrho_{1}$ and $\varrho_{2}$, we have
\begin{equation}
\mathsf{Q}_{p_{1} \varrho_1 + p_{2} \varrho_2} ( \theta ) \leq 
p_{1} \mathsf{Q}_{\varrho_1} ( \theta ) + p_{2} \mathsf{Q}_{\varrho_2} (\theta ) \, ,
\end{equation}
with $p_{1} + p_{2} = 1$. Moreover, the QFI is additive for independent measurements; that is,
\begin{equation}
\mathsf{Q}_{\varrho_1 \otimes \varrho_2} ( \theta ) =  
\mathsf{Q}_{\varrho_1} (\theta ) + \mathsf{Q}_{\varrho_2} (\theta ) \, .
\end{equation}

Second, for unitary evolutions, the QFI does not depend on the position along the orbit of $U_{\theta}$; i.e., it is the same for the state $\varrho$ as for $\varrho_{\theta} = U_{\theta}\, \varrho \, U_{\theta}^\dagger$. In such a case, the QFI simplifies to~\cite{Paris:2009aa}
\begin{equation}
\mathsf{Q}_{\varrho} ( \theta ) = 2
\sum_{i,j} \frac{(p_{i} - p_{j})^{2}}{p_{i} + p_{j}} 
| \langle i | G | j \rangle |^{2} \, ,
\end{equation}
where $p_{i}$ and $| i \rangle$ are the eigenvalues and eigenvectors of $\varrho$, respectively. For pure initial states, $\varrho = |\psi \rangle \langle \psi|$, under unitary evolution, a simpler expression for the QFI is 
\begin{equation}
\mathsf{Q}_{\psi} ( \theta ) = 4 \, \Var_{\psi} (G) 
\geq ( g_{\mathrm{max}} - g_{\mathrm{min}})^{2} \, ,
\end{equation} 
where and $g_{\mathrm{max}}$ and $g_{\mathrm{min}}$ are the maximum and minimum eigenvalues of $G$, respectively. The QFI is thus maximal for pure states that maximise the variance of $G$.

The QFI is intimately connected with the distinguishability of a probe for small variations of the parameter~\cite{Wootters:1981aa}. The distinguishability between two states, $\varrho_{1}$ and $\varrho_{2}$, can be quantified by the normalized Bures distance~\cite{Bures:1969aa} $D_{\mathrm{B}} (\varrho_{1} | \varrho_{2} ) = \sqrt{1 - F ( \varrho_{1} | \varrho_{2} )}$, where $F (\varrho_{1} | \varrho_{2} ) = \Tr [ \sqrt{ \sqrt{\varrho_{1}} \, \varrho_{2} \, \sqrt{\varrho_{1}}} ]^{2}$ is the fidelity~\cite{Uhlmann:1976aa}. Then, given two states $\varrho_{\theta}$  and $\varrho_{\theta + \delta \theta}$, obtained by an infinitesimal change in the parameter, one has~\cite{Sidhu:2020aa}
\begin{equation}
D_{\mathrm{B}}^{2} ( \varrho_{\theta} | \varrho_{\theta + \delta \theta} ) = \frac{1}{8} \mathsf{Q}_{\varrho} (\theta) \, (\delta \theta)^{2} \, ,
\end{equation} 
except for pointwise differences when $\varrho_\theta$ changes rank~\cite{Safranek:2017aa,Safranek:2018aa}. Therefore, the more distinguishable a state is, the greater the QFI and the sensitivity of the state to the parameter $\theta$.  

A final point of paramount importance is to determine  measurements that reach the ultimate precision and thus saturate the QCRB. This is equivalent to finding a POVM such that the associated FI  is equal to the corresponding QFI for  the probe state. In the single parameter case, it is always possible to saturate the QCRB by simply taking projectors onto the eigenstates of the symmetric logarithmic derivative $L_{\theta}$~\cite{Helstrom:1976ij}.  

\subsection{Multiparameter estimation}
\label{sec:qmult}

The extension to multiparameter quantum estimation looks superficially similar to the classical case. The QFI is generalized to the real-valued symmetric QFI matrix with components
\begin{equation}
[\mathsf{Q}_{\varrho} ( \btheta )]_{ij} =  \frac{1}{2} 
\Tr ( \varrho_{\btheta} \,  \{ L_{i}, L_{j} \}  ) \,, 
\end{equation} 
where $L_{i}$ is the symmetric logarithmic derivative with respect to the parameter $\theta_{i}$. For the particular case of pure states, this reduces to 
\begin{equation}
[ \mathsf{Q}_{\varrho} (\btheta) ]_{ij} =  
4 \re \, \langle \partial_{\theta_{i}} \psi_{\btheta} | 
\partial_{\theta_{j}} \psi_{\btheta} \rangle  + 4 
\langle \partial_{\theta_{i}} \psi_{\btheta} | \psi_{\btheta} \rangle 
 \langle \partial_{\theta_{j}} \psi_{\btheta} | \psi_{\btheta} \rangle 
\,, 
\end{equation} 
where $|\partial_{\theta_{i}} \psi_{\btheta}  \rangle = \partial  | \psi_{\btheta} \rangle/\partial {\theta_{i}}$. This QFI matrix verifies convexity and additivity properties~\cite{Liu:2019aa}. 

It follows that a multiparameter QCRB can then be formulated as
\begin{equation}
\bm{\Cov}_{\varrho} ( \widehat{\btheta}) \ge  \frac{1}{N} 
\bm{\mathsf{F}}_{\mathbf{x}}^{-1} ( \btheta ) \ge \, 
\frac{1}{N} 
\bm{\mathsf{Q}}_{\varrho}^{-1} (\btheta ) \, .
\end{equation}
Summing over the diagonal elements of this matrix inequality we arrive at
\begin{equation}
\sum_{i} \Var_{\varrho} (\theta_{i}) \ge {
\frac{1}{N} \Tr [ \bm{\mathsf{F}}_{\mathbf{x}}^{1} ( \btheta) ] 
\ge \frac{1}{N} \Tr [ \bm{\mathsf{Q}}_{\varrho}^{-1} (\btheta) ] \, .}
\end{equation}

For the same resources, it is now established that the simultaneous quantum estimation of multiple parameters provides better precision than estimating them individually~\cite{Baumgratz:2016aa}. Unfortunately, the possibility of attaining the ultimate quantum bounds for the simultaneous estimation is not guaranteed~\cite{Yuen:1973aa,Helstrom:1974aa,Matsumoto:2002aa,Pezze:2017aa}: the corresponding optimal measurements may not commute, thus making their implementation impossible.

The topic of multiparameter quantum estimation is nowadays quite an active field of research that has emerged as the confluence of several disparate yet interconected developments in different fields. The interested reader should consult recent reviews on the topic~\cite{Paris:2009aa,Szczykulska:2016aa,Braun:2018aa,Sidhu:2020aa,Albarelli:2020aa,Polino:2020aa,Rafal:2020aa}. For our purposes here, it is enough to mention that a sufficient condition for the joint estimation is that the operators $L_{i}$ commute.  A weaker condition is provided by the following constraint~\cite{Matsumoto:2002aa}
\begin{equation}
\label{eq:Matsu}
\Tr ( \varrho_{\btheta} 
[L_{i}, L_{j}]) = 0 \,. 
\end{equation}
For pure states, there exists a necessary and sufficient condition for the saturation of the QCRB: if $\mathsf{Q}_{\varrho} (\btheta)$ is invertible,  the QCRB can be saturated if and only if
\begin{equation}
\im \langle \psi_{\btheta} |L_{i} L_{j} | \psi_{\btheta} \rangle = 0 \, .
\end{equation}
When the condition (\ref{eq:Matsu}) is not met, there exists a tighter bound based on the so-called right logarithmic derivative~\cite{Fujiwara:1994aa}. However, this operator is not directly linked to any measurement, in contradistinction to the $L_{i}$.

\section{Rotations and the Majorana stellar representation}

Let us consider a rotation defined by its axis $\mathbf{n}= (\sin\Theta\cos\Phi,\sin\Theta\sin\Phi,\cos\Theta )^{\top}$  and angle $\theta \in [0, \pi)$, measured according to the right-hand rule. We will use the compact notation $\btheta = (\theta, \mathbf{n})$ to denote these parameters. Under this rotation, a point of the system with coordinates $\mathbf{r}$ rotates to a new position $\mathbf{r}^{\prime}$, given by
\begin{equation}
\mathbf{r}^{\prime} = \bm{\mathcal{R}}_{\btheta}  \, \mathbf{r} \,, 
\end{equation}
with
\begin{equation}
( \mathcal{R}_{\btheta})_{ij} = \delta_{ij} \cos \theta +
(1 - \cos \theta) \, n_{i} n_{j} - \varepsilon_{ijk} \, n_{k} \sin \theta \, .
\end{equation}
Henceforth, the latin indices $\{ i, j \}$ will run over the coordinate indices $\{1, 2, 3\}$, $\varepsilon_{ijk}$ is the totally antisymmetric unit vector, and summation over repeated indices is assumed. 

If one particle is in the state $| \mathbf{r}\rangle$, representing its localization at the point $\mathbf{r}$, after a rotation $\bm{\mathcal{R}}_{\btheta}$ its state must be proportional to $|\bm{\mathcal{R}}_{\btheta} \mathbf{r}\rangle$. A celebrated theorem of Wigner ensures that this action is implementable by unitary or antiunitary operators~\cite{Wigner:1931aa}. The latter possibility can be discarded because each rotation is the square of some other rotation. In consequence,
\begin{equation}
|\mathbf{r} \rangle \mathop{\longrightarrow}^{\bm{\mathcal{R}}_{\btheta}} 
R_{\btheta} | \mathbf{r} \rangle = | \bm{\mathcal{R}}_{\btheta} \mathbf{r} \rangle \, 
\end{equation}
where ${R}_{\btheta}$ is the unitary representation of the rotation (defined up to a global phase). Therefore, we have that 
\begin{equation}
\label{eq:genrot}
(R_{\btheta} \psi) (\mathbf{r}) = \langle \mathbf{r} | R_{\btheta} 
| \psi \rangle = 
\langle {\bm{\mathcal{R}}_{\btheta}}^{-1} \mathbf{r} | \psi \rangle 
\end{equation}
which shows that the wave function for a particle without internal degrees of
freedom transforms under rotations as a scalar. 

Equation (\ref{eq:genrot}) allows us to find the explicit unitary operator $R_{\btheta}$. A direct calculation shows that~\cite{Cornwell:1984aa}
\begin{equation} 
 R_{\btheta} =  \exp (- \rmi \theta \mathbf{J} \cdot \mathbf{n}) \, ,
\end{equation}
which is defined up to a $\pm$ sign due to the projective character of the mapping $\bm{\mathcal{R}} \mapsto R$ that is globally unavoidable~\cite{Galindo:1991aa}. Here, the operator $\mathbf{J}$ is the generator of all the effects of a rotation and is the total angular momentum of the system. It is an observable composed of the total orbital angular momentum $\mathbf{L}$ and the total spin $\mathbf{S}$. The three components $\{J_{i}\}$ satisfy the commutation relations of the Lie algebra $\mathfrak{su}(2)$
\begin{equation}
[J_{i}, J_{j}] = \rmi \varepsilon_{ijk} J_{k} \, .
\qquad
[\mathbf{J}^{2}, J_{i}] = 0 \, ,
\end{equation}
The irreducible representations are labelled by $J$ and spanned by the states $\{ |J m\rangle \}$, which are the simultaneous eigenstates of $\mathbf{J}^{2}$ and $J_{3}$ (with $\hbar = 1$):
\begin{equation}
\mathbf{J}^{2} |J m \rangle = J (J+1) |J m\rangle \, ,
\qquad
{J}_{z} |J m \rangle =  m |J m\rangle \, .
\end{equation}
Since $m = - J, \ldots, +J$, these states span a $(2J+1)$-dimensional Hilbert space we denote by $\mathcal{H}_{J}$. In what follows, we will assume that we work in $\mathcal{H}_{J}$, unless explicitly stated otherwise.

The notion of a Majorana representation~\cite{Majorana:1932ul} will prove to be extremely convenient for our purposes. In this representation, a pure state corresponds to a configuration of points on a sphere, a picture that makes a high dimensional Hilbert space easier to comprehend. The idea can be presented in a variety of ways~\cite{Bacry:2004aa,Bengtsson:2017aa}, but the most direct one is, perhaps, by realizing that every pure state $|\psi \rangle \in \mathcal{H}_{J}$  can be mapped onto the polynomial
\begin{equation}
  \label{eq:MajPol}
  \psi ( z) = \frac{1}{(1 + | z |^{2})^{J}}
  \sum_{m=-J}^{J} \sqrt{{2J \choose J+m}}
  \psi_{m} \, z^{J+m} \, ,
\end{equation}
where $\psi_{m} = \langle J m |\psi \rangle$ are the amplitudes of the state in the angular momentum  basis.  Up to a global factor, $| \psi \rangle$ is determined by the set $\{ z_{i} \}$ of the $2J$ complex zeros of $\psi (z)$ (suitably completed by points at infinity if the degree of $\psi (z) $ is less than $2J$).  A nice geometrical representation of $| \psi \rangle$ by $2J$ points on the unit sphere (often called the constellation) is obtained by an inverse stereographic map $\{ z_{i}\} \mapsto \{ \theta_{i}, \phi_{i} \}$, 
\begin{equation}
\label{eq:zrep}
z = \exp(\rmi \phi) \, \tan\left (\frac{\theta}{2} \right )  \,.
\end{equation}
Notice that the location of the stars has an operational meaning: a spin system, say, cannot be measured to have spin up along a direction that points away from a star on the sphere.

To illustrate how this representation works in practice, we will examine a few relevant examples. The first one is that of SU(2) or Bloch coherent states. They can be defined, among other equivalent ways, as~\cite{Perelomov:1986ly,Gazeau:2009aa} 
\begin{equation}
|z \rangle \equiv |\bm{\mathfrak{n}} \rangle = \frac{1}{(1 + | z |^{2})^{J}} 
\exp(z J_{-}) |JJ \rangle \,, 
\end{equation}
where $J_{\pm} = J_{1} \pm \rmi J_{2}$ are ladder operators, and $\bm{\mathfrak{n}}$ denotes a unit vector in the direction of spherical angles $(\theta, \phi)$ defined in (\ref{eq:zrep}). Coherent states are precisely eigenstates of the operator $\mathbf{J} \cdot \bm{\mathfrak{n}} $. For the state $|z_{0} \rangle$ we have
\begin{equation}
\psi_{z_{0}} (z) = \frac{z_{0}^{\ast}}{(1 + | z |^{2})^{J}
(1 + | z_{0} |^{2})^{J}} \left ( z + \frac{1}{z_{0}^{\ast}}\right)^{2J} 
\end{equation}
so the polynomial has a single zero at $z= - z_{0}$ with multiplicity $2J$. Consequently, the constellation collapses in this case to a single point diametrically opposed to $ \bm{\mathfrak{n}}_{0}$. 

The Majorana constellations for the angular momentum basis $\ket{Jm}$ can be easily inferred from the polynomial
\begin{equation}
\psi_{Jm} (z) = \frac{1}{(1 + | z |^{2})^{J}} {2J \choose J} 
z^{J+m}
\end{equation} 
so they consist of $J\pm m$ stars at the north and south poles, respectively. 

\begin{figure}
  \centering{\includegraphics[width=\columnwidth]{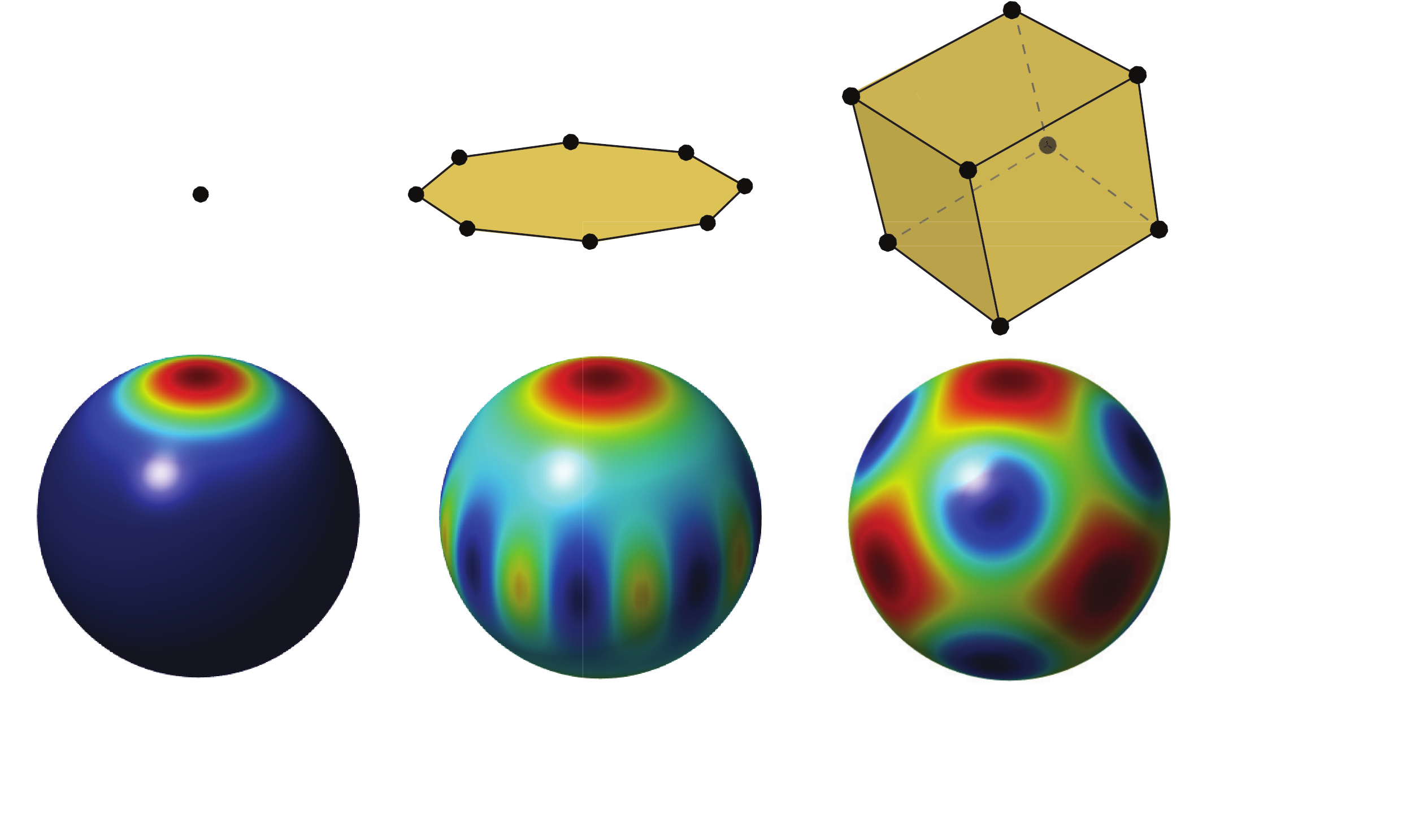}}
  \caption{{(Top) Majorana constellations for, from left to right, a Bloch coherent state, a  NOON state, and the  King of quantumness, all of them for the case  $J=4$. (Bottom) Density plots of the corresponding  Husimi functions, as defined in (\ref{eq:defHusimi}).  The colormap ranges from dark blue (corresponding to the numerical value 0) to bright red (corresponding to the numerical value 1).}}
  \label{Fig:Majorana}
\end{figure}

Another relevant set of states are the so-called NOON states, defined as~\cite{Dowling:2008aa} 
\begin{equation}
 |\mathrm{NOON} \rangle =  \frac{1}{\sqrt{2}}
  {(|JJ\rangle - |J-J\rangle)} \, ,
  \label{Eq: NOON}
\end{equation}
with associated polynomials
\begin{equation}
\psi_{\mathrm{NOON}} (z) = \frac{1}{\sqrt{2}} \frac{1}{(1 + | z |^{2})^{J}} (z^{2J} - 1)\, .
\end{equation}
The zeros are the $2J$ roots of unity, so the Majorana constellations have $2J$ stars placed around the equator of the unit sphere with equal angular separation between each star.

We can generalize the NOON states to Bloch cat states~\cite{Agarwal:1997aa}
\begin{eqnarray}
    \ket{\mathrm{cat}} & = \frac{1}{\sqrt{\mathcal{N}}} (\ket{z} - 
    \ket{- z})  \nonumber \\
       & =  \frac{1}{\sqrt{\mathcal{N}}} \sum_{m=-J}^{J}  \sqrt{{2J \choose J+m}} 
    z^m [ 1+(-1)^{J+m} ]\ket{Jm} 
\end{eqnarray} 
where $\mathcal{N} = \{ 1 - [1-|z|^{2})/(1+|z|^2]^{2J}\}/\sqrt{2}$. They interpolate from NOON states ($|z|=1$) to angular momentum eigenstates $\ket{J\pm J}$ or $\ket{J \pm (J- 1)}$ ($|z|=0,\infty$). The Majorana polynomial is
\begin{equation}
{\psi_{\mathrm{cat}} (z) = \frac{1}{\sqrt{\mathcal{N}}}  \sum_{m=-J}^{J} 
\sqrt{{2J \choose J+m}}  [z^{m} + (-z)^{m}] \, . }
\end{equation}
These again have Majorana constellations spread about the equator of the unit sphere, but now with the points spaced at azimuthal positions $\tan^{-1}\left[\cot\left(\pi k/2J\right)/z\right]$ for $k\in \{1,\cdots,2J \}$.

Since the most classical states have the most concentrated constellation, one might naively think that the most quantum states have their $2J$ stars distributed in the most symmetric fashion on the unit sphere, and this is the caase. This constitutes the realm of the  so-called Kings of Quantumness~\cite{Bjork:2015aa,Bjork:2015ab}, initially dubbed anticoherent states~\cite{Zimba:2006fk}. In a sense they are the opposite of Bloch coherent states: while the latter correspond as nearly as possible to a classical spin vector pointing in a given direction, the former \emph{point nowhere}; i.e., the average angular momentum vanishes and the fluctuations up to given order $M$ are isotropic~\cite{Hoz:2013om,Hoz:2014kq,Goldberg:2020ac}. Their symmetrical Majorana constellations herald their isotropic angular momentum properties; these states are the most sensitive for rotation measurements.  A few examples of these constellations are shown in Figure~\ref{fig:fig1}.

\section{Estimating rotation parameters}

\subsection{Known rotation axis}

We first consider estimating the angle $\theta$ by which a system is rotated about a known axis $\mathbf{n}$. Since the angular momentum operators generate rotations via
\begin{equation}
    R_{\btheta}^\dagger \, \mathbf{J} \, 
    R_{\btheta} = \bm{\mathcal{R}}_{\btheta} \, \mathbf{J} \, ,
\end{equation}
we can always take $\mathbf{n}$ as being directed along the $z$ axis. The generator is then  $G = J_{3}$, with maximal and minimal eigenvalues corresponding to eigenstates $|JJ \rangle$ and $|J-J \rangle$.  

The optimal probe states are those maximizing the variance of $J_{3}$. This requires an equal superposition of the eigenstates $|JJ \rangle$ and $|J-J \rangle$; that is, the NOON states treated in the previous section. Estimating a rotation angle for a known rotation axis is mathematically equivalent to phase estimation, explaining the ubiquity of NOON states in that context, with Heisenberg-scaling performances on average.

When the rotation axis is an arbitrary $\mathbf{n}$, the optimal probe states are the rotated versions of NOON states; i.e., $(\ket{\mathbf{n}} + \ket{- \mathbf{n}})/\sqrt{2}$, as the rotated version of the north pole $\ket{JJ}$ is precisely a Bloch coherent state. Instead of producing a different state for each axis, one may be interested in a probe state that has the best average performance for any $\mathbf{n}$. There are a few ways of quantifying such performance. First, one may consider states whose average Fisher information is largest:
\begin{equation}
    \mathsf{\bar{Q}}_{\varrho} (\theta)  = \frac{1}{4\pi} 
    \int_0^\pi {\! \! \rmd\Theta}\, \sin\Theta \, \int_0^{2\pi} \!\! \rmd\Phi\,\mathsf{Q}\left(\varrho,\theta\right) =\frac{4}{3} \sum_{i=1}^3\Var_{\varrho_\theta} (J_i).
\end{equation} 
Optimizing this quantity is the same as optimizing the signal-to-noise ratio for a given total angular momentum $J$. The minimum is achieved by states whose angular momentum projection is maximal in any {fixed} direction; i.e., they satisfy $\Tr(\varrho_\theta \mathbf{J}) = J \mathbf{n}^\prime$. For pure states, they are $\ket{\mathbf{n}^\prime}$  and $\mathsf{\bar{Q}}_{\mathrm{min}}=J$. The maximum is achieved by states whose angular momentum projection vanishes in all directions; i.e., $\Tr (\varrho_\theta \mathbf{J} )=\mathbf{0}$, with $\mathsf{\bar{Q}}_{\mathrm{max}} =\case{4}{3}J(J+1)$. In this sense, the most sensitive probe states are those whose classical angular momentum features are hidden (they include the NOON states).

One can also find states that optimize the average variance of the estimated parameter by averaging over the QCRB (\ref{eq:QCRB}):
\begin{equation}
    \overline{\Var}_{\varrho} (\theta) = 
    \frac{1}{4\pi} \int_0^\pi \!\! \rmd \Theta\,\sin\Theta 
    \int_0^{2\pi}\!\! \rmd \Phi \,\frac{1}{\mathsf{Q}_{\varrho} (\theta)} \, .
 \end{equation} 
 In that case, the average over all angles diverges for the states $R_{\btheta^\prime}\ket{Jm}$ for any $\btheta^\prime$,\footnote{Incidentally, this supports the conjecture in Ref.~\cite{Goldberg:2020aa} that the average over all angles diverges for more than just the states $\ket{\bm{\mathfrak{n}}}$, but we have not proven the nonexistence of other such states.} while it is minimized {by states satisfying}
\begin{equation}
    \Tr ( \varrho_\theta \, \mathbf{J} )=\mathbf{0} \, , 
    \qquad
    \Tr (\varrho_\theta \, \{J_i,J_j \} ) =  
    \frac{2}{3}J(J+1) \delta_{ij} \, ,
\end{equation} 
with value $\case{3}{4} J(J+1)$. The states achieving the minimal average variance are those whose first and second order angular momentum properties are isotropic, which are known as Kings of Quantumness. These have Heisenberg-scaling uncertainties on average. NOON states, as an intermediate, achieve an average variance of $\arctan(\sqrt{2J-1})/(2J\sqrt{2J-1})$, which scale more poorly with $J$ than the Kings but better than the states $\ket{\mathbf{n}^\prime}$. This shows how the problem of rotation estimation requires more nuance than that of phase estimation~\cite{Goldberg:2020aa}. The outstanding performance of estimating rotation angles about various axes has been demonstrated using light's orbital angular momentum degree of freedom in Ref.~\cite{Bouchard:2017aa}.

{It is worth mentioning that the Kings of Quantumness also have isotropic higher-order moments. Since the only criterion generally accepted for assessing the optimality of an estimator is whether its variance saturates the QCRB, the standard approach might overlook relevant  information that may be present in the complete parameter distribution, and that can be captured by looking at higher-order moments~\cite{Gianani:2020aa}.}

\subsection{Unknown rotation axis}

The rotation axis $\mathbf{n}$ is not always known \emph{a priori}. Then, we must consider the multiparameter estimation problem of determining all three components of $\btheta$. We mainly focus on our $\btheta$ parametrization due to its physical importance, but our results are applicable to the various parametrizations one can find in the literature. Of note, one can always find the QFI matrix for a different parametrization by using equation~(\ref{eq:FI change of param}).

Before determining and optimizing the QFI matrix for this multiparameter scenario, we mention that an alternative approach to optimizing states for probing rotations about arbitrary axes is that of optimal quantum rotosensors~\cite{Chryssomalakos:2017aa,Martin:2020aa}. These are probe states that are optimally sensitive to determining whether a system has undergone a rotation by known angle $\theta$ about unknown axis $\mathbf{n}$~\cite{Martin:2020aa}.  This is done by finding states that are, on average, the most changed by a rotation for a fixed angle. Then, the most sensitive states for identifying the presence of a small rotation angle are the Kings, of a large rotation angle are the states $\ket{\mathbf{n}^\prime}$, and of intermediate rotation angles are other states. Optimal quantum rotosensors can be used to determine whether or not a rotation is present; if present, one can use the results presented here to optimally determine the parameters of the rotation.

The three rotation parameters being estimated correspond to three generators defined as $G_k = (\rmi \partial_{\theta_k} R_{\btheta}) R^\dagger_{\btheta}$.  Computing these generators is nuanced because $[\partial_{\theta_k} (\mathbf{J}\cdot\mathbf{n}), \mathbf{J}\cdot\mathbf{n}]\neq 0$. To fix the problem, we express the generators as~\cite{Wilcox:1967aa}
\begin{equation}
    G_{k} = \left (\int_0^1 d\alpha \, 
    \rme^{-\rmi\alpha \mathbf{J} \cdot \bomega}
    \mathbf{J} \rme^{\rmi\alpha \mathbf{J}\cdot \bomega} \right) \cdot 
    \partial_{\theta_k} \bomega \equiv 
    \mathbf{J}\cdot\mathbf{g}_k(\btheta),
    \label{eq:rotation generators}
\end{equation} 
where $\bomega= \theta\mathbf{n}$, and the final expression is guaranteed to be a linear combination of angular momentum operators by the Baker-Campbell-Hausdorff formula~\cite{Suzuki:1985aa} or equivalently by the properties of rotation operators. This allows us to compute the vectors $\mathbf{g}_k(\btheta)$ using any representation of angular momentum, for example by computing the vectors using Pauli matrices as done in Ref.~\cite{Hou:2020aa}. We find the vectors $\mathbf{g}_k(\btheta)$ to be given by
\begin{eqnarray}
    \mathbf{g}_\theta & = &  \mathbf{n} = \left ( 
    \begin{array}{c}
    \cos \Phi \sin \Theta\\
    \sin \Phi \sin \Theta \\
    \cos \Theta
    \end{array} 
    \right ), \nonumber \\
     \mathbf{g}_\Theta  & = & 2\sin \left ( \frac{\theta}{2} \right ) \left [\cos\left ( \frac{\theta}{2} \right ) \, 
     \frac{\partial \mathbf{n}}{\partial \Theta} - 
     \sin\left ( \frac{\theta}{2} \right ) \, \frac{\partial \mathbf{n}}{\partial \Theta} \times \mathbf{n}\right ] \, ,  \\
    \mathbf{g}_\Phi & = & 
    2\sin\left ( \frac{\theta}{2} \right ) 
    \left [ \cos\left ( \frac{\theta}{2} \right ) \; 
    \frac{\partial \mathbf{n}}{\partial \Phi} -
    \sin \left ( \frac{\theta}{2} \right ) \, 
    \frac{\partial \mathbf{n}}{\partial \Phi} \times \mathbf{n}\right ]  
    \, . \nonumber
\end{eqnarray}
Then, the QFI matrix can be expressed as
\begin{equation}
   \bm{\mathsf{Q}}_{\psi} ( \btheta ) = 4 
   \mathbf{G}^\top (\btheta) \,  \bm{\mathsf{C}}(\psi_{\btheta} ) 
   \, \mathbf{G}(\btheta) \, .
\end{equation}
$\mathbf{G} (\btheta ) = \left(\mathbf{g}_\theta \quad \mathbf{g}_\Theta \quad \mathbf{g}_\Phi \right)^{\top}$ encompasses the angular information of the three parameters and 
\begin{equation}
    \mathsf{C}_{ij} (\psi_{\btheta} ) = 
    \Cov_{\psi_{\btheta}} (J_i,J_j )
\end{equation} 
the sensitivity characteristics of the rotated state.  Since the covariances may vary with $\btheta$, we can consider the transformation property
\begin{equation}
    \bm{\mathsf{C}} (\psi_{\btheta} ) =
    \bm{\mathcal{R}}^{\top}_{\btheta} \; \bm{\mathsf{C}} (\psi ) \; 
    \bm{\mathcal{R}}_{\btheta}
\end{equation} 
and incorporate the rotation matrices into the angular matrices
\begin{equation}
    \bm{\mathsf{Q}}_{\psi} (\btheta) = 
    4 \tilde{\mathbf{G}}^\top (\btheta) \,
    \bm{\mathsf{C}} (\psi )\,
    \tilde{\mathbf{G}} (\btheta)  \,,
    \label{eq:rotations QFIM btheta}
\end{equation}
with $\tilde{\mathbf{G}} (\btheta ) = \bm{\mathcal{R}}_{\btheta} \, \mathbf{G} (\btheta)$. Equation~(\ref{eq:rotations QFIM btheta}) has the same form as the  change-of-parametrization formula for the QFI matrix in equation~(\ref{eq:FI change of param}). One can easily switch to a different parametrization: the Jacobian governing this change will, similar to $\tilde{\mathbf{G}} (\btheta )$, only depend on geometric properties of the coordinate systems, and so the final QFI matrix will always have the form $\bm{\mathsf{Q}}_{\psi} ( \bm{\vartheta} ) =   4 \bm{\mathcal{G}}^\top (\bm{\vartheta}) \,  \bm{\mathsf{C}} (\psi) \bm{\mathcal{G}}(\bm{\vartheta})$ for any set of parameters $\bm{\vartheta}$. The crucial point is that in order to find states optimally suited for estimating arbitrary unknown rotations one must optimize the \emph{sensitivity covariance matrix} $\bm{\mathsf{C}} (\psi)$.

First, we identify when the QFI matrix can and cannot be inverted. A singular QFI matrix implies that the triad of parameters cannot be estimated for a given state and a given parametrization (see \ref{app:Singular QFIM} for further discussion of singular QFI marices). This is the case when either $\bm{\mathsf{C}}(\psi)$ is singular, implying that the state will never be useful for estimating all of the parameters of a rotation, or when $\bm{\mathcal{G}}$ is singular, implying that the coordinate system is singular at that specific set of parameters regardless of the probe state. As discussed in Ref.~\cite{Goldberg:2018aa}, singularities in one coordinate system can be alleviated for specific parameters by switching to a new coordinate system. For example, the spherical and $ZYZ$ Euler angle parametrizations are singular for a rotation by $0$, but the Cartesian and $XYZ$ Euler angle parametrizations are invertible there, and the Cartesian parametrization is singular for rotations by $2\pi$ and the $XYZ$ Euler angle parametrization when the $Y$ rotation is by $\pm \pi/2$. 

It is straightforward to show that $\bm{\mathsf{C}} (\psi )$ is singular if and only if the probe state is an eigenstate of some angular momentum projection; that is, proportional to $R_{\btheta^\prime}\ket{Jm}$. Similar to the single-parameter scenario, where eigenstates of the angular momentum projection operators were the least useful for estimating rotations about unknown axes, we see that states with any definite angular momentum projection cannot be used for simultaneously estimating all three parameters of a rotation.

To find the most sensitive states we seek to minimize the trace of the inverse of $\bm{\mathsf{C}} (\psi )$. This is straightforward to optimize because, for any symmetric, invertible matrix $\mathbf{M}$, $\Tr ( \mathbf{M} \mathbf{M}^{-1})^2 \leq \Tr (\mathbf{M}) \Tr (\mathbf{M}^{-1})$, with equality if and only if $\mathbf{M}$ is proportional to the identity matrix. Since  $\Tr [\bm{\mathsf{C}} (\psi) ]=J(J+1)$ is fixed by the Casimir invariant, we find
\begin{equation}
    \Tr [\bm{\mathsf{C}}^{-1}\left(\psi\right) ]\geq\frac{9}{J(J+1)},
\end{equation} 
with the trace of the inverse achieving the minimum only for the Kings of Quantumness. We again see that having isotropic angular momentum up until second order makes a state most sensitive to arbitrary rotations about arbitrary axes. Next we showcase how to use these states to saturate the QCRB for a particular parametrization.

\subsection{Optimal measurements}

Saturability of the QCRB hinges upon the expectation values of the commutators of the generators defined by equation~(\ref{eq:rotation generators}). Fortunately, for states with isotropic covariance matrices, the expectation values of the commutators are guaranteed to vanish. In fact, these expectation values will vanish for all states whose average angular momentum vanishes, in any parametrization with generators defined by $G_k=\mathbf{J}\cdot\mathbf{g}_k$. The Kings thus guarantee that all three parameters can be simultaneously estimated at a precision saturating the QCRB for any triad of rotation parameters.

We can find a projection-valued measure (PVM) that saturates the QCRB by considering projections onto the three states proportional to $G_k\ket{\psi_{\btheta}}$ and onto the original rotated state $\ket{\psi_{\btheta}}$. Because the vectors $\mathbf{g}_k$ are orthogonal, these four states are all mutually orthogonal for the Kings. A PVM consisting of the state $\ket{\psi_{\btheta}}$ and any orthonormal combination of the states $G_k\ket{\psi_{\btheta}}$ suffices to produce a classical {FI} matrix that equals the QFI matrix. When the probe states only have their first-order angular momentum projections vanish, one must use a Gram-Schmidt orthogonalization procedure among the states $G_k\ket{\psi_{\btheta}}$~\cite{Pezze:2017aa}. 

The QCRB can be saturated in the asymptotic limit, when $\widehat{\btheta}$ tends to the true value $\btheta$. In this case, one has a good estimate $\ket{\psi_{\widehat{\btheta}}}$ that differs from the true state by some small, unknown rotation angle $\varepsilon\ll 1$ about some unknown axis $\mathbf{u}$:
\begin{equation}
\ket{\psi_{\btheta}}= \exp(\rmi \varepsilon \mathbf{J}\cdot\mathbf{u}) 
\ket{\psi_{\widehat{\btheta}}}  \simeq \left [ 1+ \rmi \varepsilon  \mathbf{J}\cdot\mathbf{u} - \case{1}{2} \varepsilon^2 (\mathbf{J}\cdot\mathbf{u})^2 \right]
\ket{\psi_{\widehat{\btheta}}}.
\end{equation}
Defining four orthonormal projectors $\Pi_k=\ket{\phi_k}\bra{\phi_k}$ by the states
\begin{equation}
    \ket{\phi_0}= \ket{\psi_{\widehat{\btheta}}}, \qquad 
    \ket{\phi_k}=\frac{\mathbf{v}_{k} \cdot \mathbf{J}}{\sqrt{J(J+1)\mathbf{v}_{k} \cdot \mathbf{v_k}}}\ket{\psi_{\widehat{\btheta}}}, 
\end{equation}
where we can choose the three vectors $\mathbf{v}_{k}$ to be orthonormal, we find the expectation values of the projectors to  be
\begin{equation}
\fl   p_0\equiv \langle{\Pi_0} \rangle \simeq 1- \case{1}{3} {\varepsilon^2} J(J+1), 
\qquad 
p_j\equiv \langle {\Pi_j} \rangle \simeq \case{1}{3} {\varepsilon^2} J(J+1) 
(\mathbf{v}_k\cdot \mathbf{u})^2 \,.
\end{equation}
In this limit, that the probability $p_0$ approaches unity and the probabilities $p_k$ are vanishing shows that the estimates $\widehat{\btheta}$ are approaching the true values $\btheta$. In addition, we have:
\begin{eqnarray}
    \partial_i p_k&= &\frac{3\rmi}{J(J+1)} \frac{\bra{\phi_0} 
    (\mathbf{J}\cdot\mathbf{v}_k ) (\mathbf{J}\cdot\mathbf{g}_i )
     \ket{\psi_{\btheta}} \bra{\psi_{\btheta}} \mathbf{J}\cdot\mathbf{v}_k \ket{\phi_0}}{\bra{\phi_0} (\mathbf{J}\cdot\mathbf{v}_k)^2\ket{\phi_{0}}} +\mathrm{c.c.} \nonumber \\
        &= &\case{2}{3} {\varepsilon} J(J+1) (\mathbf{v}_k\cdot \mathbf{g}_i ) 
        (\mathbf{v}_k\cdot\mathbf{u} )   \,,
\end{eqnarray} 
which holds for any state whose angular momentum projection vanishes and the final equality is valid for any state whose second-order angular momentum features are isotropic. From this we can immediately compute (the derivatives of $p_0$ vanishing to lowest order in $\varepsilon$)
 \begin{equation}
    \bm{\mathsf{F}}_{k} ( \btheta) = \case{4}{3} J(J+1) 
    \mathbf{G}^\top (\btheta) \mathbf{G}(\btheta ) = 
    \bm{\mathsf{Q}}_{\varrho} ( \btheta)\, ,
\end{equation}
where $G_{ij} (\btheta )= \mathbf{v}_k \cdot \mathbf{g}_{j}$. A PVM comprised of, say, the state $\ket{\psi_{\widehat{\btheta}}}$ and $J_k\ket{\psi_{\widehat{\btheta}}}$ for $k\in (1,2,3)$ will thus saturate the QCRB for optimal states. 

{Thus far, we have been considering only ideal scenarios. The fundamental bounds change in the  presence of losses and decoherence.  Actually, NOON states are extremely susceptible to losses: in the photonic case, a loss of, e.g., a single photon makes the state completely useless for sensing. The Kings of Quantumness behave much better in the presence of losses.}

{In practice, however, those optimal states are notoriously hard to prepare, apart from the regime of very small $J$. For large photon numbers, the only practically accessible states of light are squeezed states~\cite{Andersen:2016aa}. We will see in Sec.~\ref{sec:Metpow} why these states are indeed optimal in the asymptotic limit of very large $J$.}

\subsection{Suboptimal measurements}

The PVM described in the above section may be challenging to implement experimentally. Easier is to project the rotated state onto a set of {Bloch coherent states} for various directions and to reconstruct the rotation parameters from these measurements. 

The {set of projections}
\begin{equation}
\label{eq:defHusimi}
    {\mathfrak{q}_{\bm{\mathfrak{n}}} = | \langle \bm{\mathfrak{n}} |\psi_{\btheta} \rangle|^2 }
\end{equation} 
{constitute the Husimi function}~\cite{Husimi:1940aa}. Knowledge of all of the projections ${\mathfrak{q}_{\bm{\mathfrak{n}}}}$ is tantamount to knowledge of the rotated state $\ket{\psi_{\btheta}}$, but such information is redundant: {it suffices to sample the function at a few locations $q_{\bm{\mathfrak{n}}}$ and use these results to orient the Husimi function and thus estimate the rotation parameters.} This method can be used for any probe state {and amounts to a positive operator-valued measure (POVM) rather than a PVM because the states $\ket{\bm{\mathfrak{n}}}$ are not mutually orthogonal for differing $\bm{\mathfrak{n}}$.}

At how many locations must the Husimi function be sampled to uniquely orient it? In general, the answer depends on the probe state in question and the locations being sampled, but we argue using concepts applied from geographical positioning systems that four projections $q_{\bm{\mathfrak{n}}}$ should suffice for this orientation problem. 

Projecting the rotated state $\ket{\psi_{\btheta}}$ onto an arbitrary {Bloch coherent state $\ket{\bm{\mathfrak{n}}_{1}}$ amounts to sampling the Husimi function at $\bm{\mathfrak{n}}_{1}$.} The value of this first projection {$\mathfrak{q}_{\bm{\mathfrak{n}}_{1}}$} defines a set of \textit{level curves}, and the state \textit{must} be oriented in such a way that $\mathbf{n_1}$ lies on one of these curves. Rotating the state along any of these level curves will produce the same value $\mathfrak{q}_{\bm{\mathfrak{n}}_{1}}$.

Projecting the rotated state onto another coherent state {$\ket{\bm{\mathfrak{n}}_{2}}$} defines another set of level curves. Rotating the state along these level curves again produces the same value {$\mathfrak{q}_{\bm{\mathfrak{n}}_{2}}$}, so in general we expect there to be multiple \textit{intersection points} for orienting the {Husimi} function such that $\bm{\mathfrak{n}}_{1}$ lies along a curve $\mathfrak{q}_{\bm{\mathfrak{n}}_{1}}$ and $\bm{\mathfrak{n}}_{2}$ lies along a curve $\mathfrak{q}_{\bm{\mathfrak{n}}_{1}}$.

Barring pathological cases, a third projection uniquely specifies one of the above intersection points for orienting the {Husimi} function. A fourth projection helps deal with pathological cases, and a fifth projection helps with normalization. Viewed in a different way, the set of $d$ angular coordinates $\bm{\mathfrak{n}}_i$ can be rigidly rotated until the $d$ projections {$\mathfrak{q}_{\bm{\mathfrak{n}}_{i}}$} match the given state $\ket{\psi_{\btheta}}$. The obvious pathological cases are those for which the {Husimi} functions at a set of $d$ angular coordinates are equal to the {Husimi} functions at a rigid rotation of those angular coordinates; this is easily mitigated by knowing the nonoriented {Husimi} function ahead of time and thus knowing its pathologies.

\begin{figure}
    \centering
    \includegraphics[width=0.90\columnwidth]{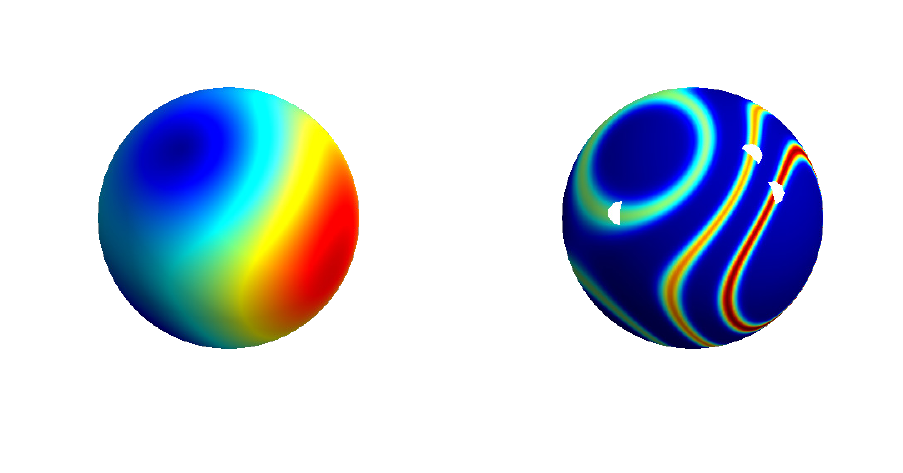}
    \caption{(Left) Husimi function for probe state with randomly-chosen components in the $\ket{2m}$ basis, scaled as $(4\mathfrak{q}_{\bm{\mathfrak{n}}}/\pi)^{3/4}$. Orienting this distribution is equivalent to estimating the parameters of a rotation. (Right) Level curves of the Husimi function at three fiducial values $\mathfrak{q}_{\bm{\mathfrak{n}}}=1/10,\,3/10,\,5/10$, each with variance $1/20000$, which are ``measured'' to be at polar angles $1, 0.6, 0.9$ and azimuthal angles $-1.7, -0.3, -0.3$, respectively (the white semicircles). The distribution is rotated from the one on the left such that the three measured points coincide with the value of the Husimi function at those points.}
    \label{fig:fig3}
\end{figure}

Any set of projections that avoids pathological cases will suffice for discerning the rotation parameters. Uncertainty in the values {$\{ \mathfrak{q}_{\bm{\mathfrak{n}}} \}$} leads to ``fuzziness'' that broadens the level curves for each projection, so increasing the number of projections increases the overall precision, just like increasing the number of satellites measuring distances increases precision for global positioning systems. The limit of measuring projections in all directions is tantamount to performing full quantum state tomography on the rotated probe state $\ket{\psi_{\btheta}}$.
As an example, consider the state in Figure~\ref{fig:fig3}, which has the randomly-chosen components $\{0.000394688\, +0.409134 \rmi,0.0324599\, +0.0448131 \rmi,0.494021\, +0.484609 \rmi,0.483644\, +0.114779 \rmi,0.100279\, +0.305783 \rmi\}$ in the $\ket{2m}$ basis. The figure depicts a scaled version of {Husimi} function for all values of $\bm{\mathfrak{n}}$. We can consider various level curves of this function, with their uncertainties, by plotting Gaussians centered at particular values $\mathfrak{q}_{\bm{\mathfrak{n}}}$. For a certain rotation $R_{\btheta}\ket{\psi}$, three separate measurements $q_{\bm{\mathfrak{n}}_i}$ at three values of $\bm{\mathfrak{n}}_i$, the white semicircles in Figure~\ref{fig:Q}, produce values $\mathfrak{q}_{\bm{\mathfrak{n}}_{1}}=1/10$, $\mathfrak{q}_{\bm{\mathfrak{n}}_{2}}=3/10$, and $\mathfrak{q}_{\bm{\mathfrak{n}}_{3}}=5/10$. This tells us that the {Husimi} function \textit{must} be oriented in such a way that the three directions lie on the three level curves.

The projections $\{ \mathfrak{q}_{\bm{\mathfrak{n}}} \}$ can be used in, for example, maximum likelihood estimation schemes to determine the angular parameters $\btheta$. {Actually, we can sketch the usefulness of this POVM in the language of Fisher information. To that end, we assume that the sampled directions are enough to provide a (approximate) resolution of the identity for the rotated probe state:}
\begin{equation}
    {\sum_{\bm{\mathfrak{n}}} \mathfrak{q}_{\bm{\mathfrak{n}}}= 
    \bra{\psi_{\btheta}}\left( \sum_{\bm{\mathfrak{n}}} \ket{\bm{\mathfrak{n}}}\bra{\bm{\mathfrak{n}}}\right) 
    \ket{\psi_{\btheta}}= \openone \, .}
\end{equation}
{The classical Fisher information from the projections $\mathfrak{q}_{\bm{\mathfrak{n}}}$  onto Bloch coherent states takes the form}
\begin{eqnarray}
    {[\mathsf{F}_{\bm{\mathfrak{n}}} (\btheta ) ]_{ij} } & = &
    4\left(\bra{\psi_{\btheta}}\frac{\left\{\mathbf{J}\cdot\mathbf{g}_i,\mathbf{J}\cdot\mathbf{g}_j\right\}}{2}\ket{\psi_{\btheta}}
    -
    \bra{\psi_{\btheta}}\mathbf{J}\cdot\mathbf{g}_i\ket{\psi_{\btheta}}\bra{\psi_{\btheta}}\mathbf{J}\cdot\mathbf{g}_j\ket{\psi_{\btheta}}\right) \nonumber \\
    & + & 2 \int d\bm{\mathfrak{n}} \re \left ( 
     \frac{\langle \psi_{\btheta} | \mathbf{J}\cdot\mathbf{g}_i | \bm{\mathfrak{n}} \rangle}
     {\langle \psi_{\btheta} |\bm{\mathfrak{n}} \rangle} \right ) \im\left(\langle \psi_{\btheta} | \bm{\mathfrak{n}}\rangle\langle\bm{\mathfrak{n}}| L_j |\psi_{\btheta}\rangle\right) \, ,
     \label{eq:GPS Fisher}
\end{eqnarray} 
{where we have averaged over Bloch coherent states. In this limit, measuring a sufficient number of projections lets one approach the QCRB, regardless of probe state, and a necessary condition for saturating the QCRB is the vanishing of the imaginary term in (\ref{eq:GPS Fisher}).}

\section{Metrological power of Majorana constellations}
\label{sec:Metpow}

The geometrical structure of a quantum state informs its usefulness for rotation sensing. This is because rotations acting on a quantum state rigidly rotate the latter's Majorana constellation, making the rotation sensing problem equivalent to the distinguishing between constellations' orientations. We can see geometrically why some states are not useful for general rotation sensing, some states are useful for estimating rotations about known axes as in phase estimation, and others are useful for estimating rotations about unknown axes.

{ We begin with SU(2) coherent states.  Their constellations only possess two angular degrees of freedom, because they have a continuous rotational symmetry about one axis, so can never be used for sensing all three parameters of a rotation. This is reflected by the properties of the sensitivity covariance matrix {(which we write for the particular case of a coherent state located at the north pole)}
\begin{equation}
    \bm{\mathsf{C}} (\ket{JJ} ) = 
    \left(\begin{array}{ccc}
        \frac{J}{2} & 0&0\\0&\frac{J}{2}&0\\0&0&0
    \end{array}\right) \,.
\end{equation} 
The matrix has determinant zero, its condition number is formally infinite, and the trace of its inverse diverges.

Next, we turn to eigenstates of angular momentum projection operators.  Again, the associated constellations only have two angular degrees of freedom, which is why it cannot be used to simultaneously estimate all three rotation parameters. The sensitivity covariance matrix
\begin{equation}
    \bm{\mathsf{C}} (\ket{Jm} ) = 
    \left(\begin{array}{ccc}
        \case{1}{2} (J^2+J-m^2) & 0 & 0\\
        0 & \case{1}{2} (J^2+J-m^2) & 0 \\
        0 & 0 & 0
    \end{array}
    \right) 
\end{equation} 
seems more useful than $\bm{\mathsf{C}} (\ket{JJ} )$, but it still cannot be inverted.

It is well known that NOON states are useful for estimating rotations about known axes.  Their constellations have an extra angular degree of freedom that yields a sensitivity covariance matrix
\begin{equation}
    \bm{\mathsf{C}}(| \mathrm{NOON} \rangle ) = 
    \left(\begin{array}{ccc}
        \frac{J}{2} & 0&0\\
        0 &\frac{J}{2}&0\\
        0 & 0 & J^2
    \end{array}
    \right) \, .
\end{equation} 
Now we see that the matrix has a nonzero determinant $J^4/4$, its condition number is $J/2$, and the trace of its inverse is $\left(4J+1\right)/J^2$. These states are useful for estimating rotations around a particular axis, evidenced by their geometrical representation and the $\mathsf{C}_{33}$ component, but their performance for arbitrary rotations gets poorer with increasing $J$ (as seen by the increasing condition number) and leads to a total uncertainty that only scales as $\mathcal{O}\left(1/J\right)$.

Bloch cat states also have a diagonal sensitivity covariance matrix, just like for NOON states. Still, the determinant decreases monotonically and the trace of the inverse increases monotonically with $|\Theta-\pi/2|$, so we see that these Bloch cat states monotonically span the transition between the performance of NOON states and the inferior angular momentum eigenstates.

Majorana constellations that extend beyond a single great circle to make use of the three-dimensional structure of the unit sphere are instrumental to optimizing estimation of rotations about arbitrary axes and of all three rotation parameters. We can use this geometrical understanding to concoct other useful states for metrology. For example, to improve the usefulness of {NOON} states we can consider supplementing the stars on the equator of the unit sphere with some stars at the north and south poles:
\begin{equation}
    \ket{\psi_{\mathrm{balanced}}} = {\frac{1}{\sqrt{2}} 
    (\ket{Jm} + \ket{J-m} )  \, .}
\end{equation}
These now have $2m$ points spread about the equator of the unit sphere, supplemented by $J-m$ points at each of the north and south poles (we take $m>1/2$ without loss of generality to ensure that the states have vanishing angular momentum projections). How can we determine the optimal number of stars to put at the poles and along the equator? We know that these states will again be useful for estimating rotations about arbitrary axes, with most of their usefulness being for rotations about a single axis, when $m \sim J$. We can determine the optimal balance through the sensitivity covariance matrix:
\begin{equation}
    \bm{\mathsf{C}} (\ket{\psi_{\mathrm{balanced}}} )= 
    \left(\begin{array}{ccc}
        \case{1}{2} (J^2+J-m^2) & 0&0\\0& 
        \case{1}{2} (J^2+J-m^2)&0\\0 & 0 & m^2
    \end{array}\right) \, .
\end{equation} 
We see that decreasing $m$ from $J$ while retaining $m=\mathcal{O}(J)$
makes the matrix more stable to inversion and the total uncertainty decrease, relative to NOON states. The optimal value of $m^2= \case{1}{3} J(J+1)$ yields a sensitivity covariance matrix $\bm{\mathsf{C}} (\ket{\psi_{\mathrm{balanced}}} ) \to \case{1}{3} J(J+1) \openone$, the same as for the Kings; this simultaneously maximizes the determinant, minimizes the condition number, and minimizes the trace of the inverse.
This optimal value of $m$ can only be obtained for some values of $J$ [$2J\in\left(6, 25, 96, 361, 1350, 5041,18816, 70225, 262086, 978121\cdots\right)$], making those states Kings, and for other values of $J$ these states can only approach the Kings. This geometrical approach shows how rotation sensing performance can be improved by adding three-dimensional structure to the Majorana constellation.

The states with the best geometries are the Kings of Quantumness. These have Majorana constellations with discrete symmetries about multiple axes, whose highly distributed stars yield the sensitivity covariance matrix
\begin{equation}
    \bm{\mathsf{C}}(\ket{\psi_{\mathrm{Kings}}} ) = 
    \frac{1}{3} J(J+1) \left(\begin{array}{ccc}
        1 & 0&0\\0&1&0\\0&0&1
    \end{array}\right) \, .
\end{equation} This has a determinant that scales with $\mathcal{O}\left(J^6\right)$, a perfect condition number of 1, and the trace of the inverse scales with $\mathcal{O} (1/J^2 )$. From all of these perspectives it is clear that they have more metrological power for estimating arbitrary rotations than any of the other states considered in the context of phase estimation.

As an extension to these states, we can consider states that do not live in a single Hilbert space $\mathcal{H}_J$, as such states have been used in phase estimation, and compare their performances to their geometrical representations. The Majorana representation must be extended to a set of nested spheres, one for each $\mathcal{H}_J$, and the information about relative phases between a state's components in each Hilbert space is lost. {To make the connection to angular momentum, we represent each state by its occupation of two harmonic oscillator modes annihilated by $a$ and $b$, and use the Schwinger} map~\cite{Schwinger:1965kx} 
\begin{equation}
  \label{eq:1}
  {J}_{\mu} = \case{1}{2}
  ( 
  \begin{array}{cc} 
    {a}^{\dagger} & {b}^{\dagger}
  \end{array}
  ) \,
  \sigma_{\mu} \,
  \left (
  \begin{array}{c}
    {a} \\
    {b}
  \end{array} 
  \right ) \, .
\end{equation}
The Greek index $\mu$ runs from 0 to 3, with $\sigma_{0} = \openone$ and $\{
\sigma_{k} \} $ ($k = 1, 2, 3$) are the Pauli matrices. Note carefully that ${J} = {N}/2$, where ${N} = {a}^{\dagger} {a} + {b}^{\dagger} {b}$ is the operator for the total number of excitations. 

\begin{figure}
    \centering
    \includegraphics[width=0.75\columnwidth]{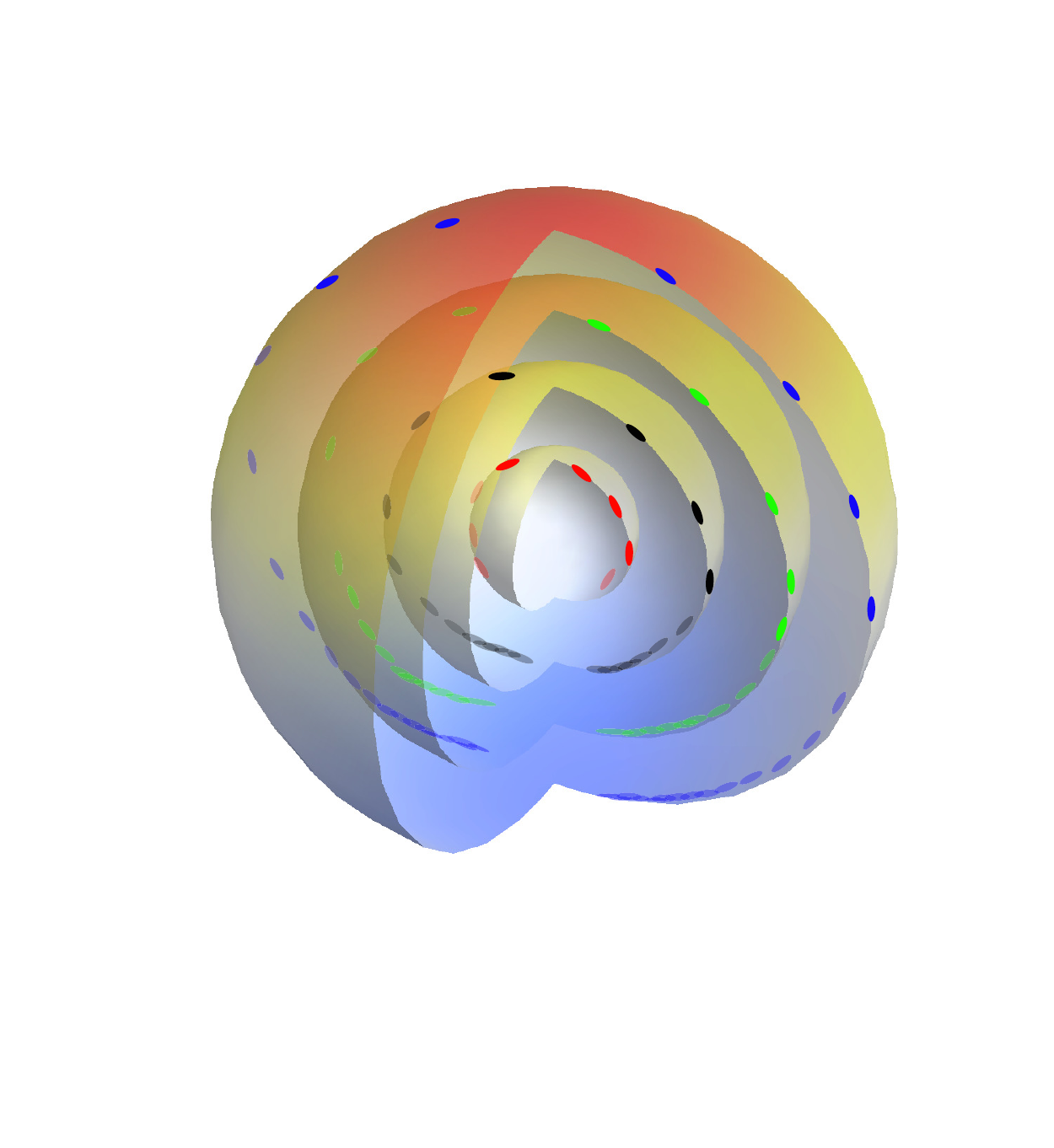}
    \caption{Majorana constellations in the photon-number subspaces with $N=(4,9,14,19)$ for two-mode photonic states with a coherent state in one mode and a squeezed state in the other. All of the constellations in each photon-number subspace lie on the same longitudinal great circle. The ratio of the coherent state strength to the squeezing parameter is $\alpha^2/\lambda=4$ and a change in its phase merely rotates the constellation. The most sensitive states have $|\lambda|\approx 1$; this is reflected in the constellations being the most evenly distributed for photon-number subspaces near $|\alpha^2/\lambda|$.}
    \label{fig:Q}
\end{figure}

For example, we can begin with the two-mode coherent states 
\begin{equation}
    \ket{\alpha,\beta}\propto \exp\left(\alpha a^\dagger+\beta b^\dagger\right)\ket{\mathrm{vac}} \, ,
\end{equation}
{where $|\mathrm{vac}\rangle$ denotes the two-mode vacuum.} In each subspace $\mathcal{H}_J$, this state is represented by $2J$ stars at the same angular location: the polar angle is always $2\tan^{-1} |\beta/\alpha |$ and the azimuthal angle is always $\arg(\beta/\alpha)$. These states have
\begin{equation}
    \bm{\mathsf{C}} (\ket{\alpha,\beta}) = \frac{|\alpha |^2 + |\beta |^2}{4}
    \left(\begin{array}{ccc}
        1 & 0 & 0\\
        0 & 1 & 0\\
        0 & 0 & 1
    \end{array}
    \right) \, .
\end{equation} 
Similar to the Kings, the condition number is 1, making these states stable to matrix inversion. However, for average energies $|\alpha |^2 + |\beta |^2$ and $2J$, the two-mode coherent states scale much more poorly with energy in the context of determinant and total uncertainty. This can be interpreted from the geometry: the coherent states do not possess much angular information, because all of their stars lie on the same axis, but the hidden relative phase information can be used to determine rotations around arbitrary axis with \textit{some} precision, because the states are not eigenstates of an angular momentum projection operator.

The previous example is readily extended to other states useful for phase estimation. It can be shown that all of the stars in all of the subspaces lie along the same great circle for many such states, explaining why these are sensitive to estimating rotations about particular axes. For example, the two-mode squeezed states have half of their stars lying at opposite poles from each other in each subspace, giving more angular resolution than coherent states alone. A more intricate case is that of a state that is coherent in one mode and squeezed in the other: 
\begin{equation}
    \ket{\psi_{\mathrm{c+s}}} \propto \exp\left( \alpha a^\dagger + 
    {\case{1}{2} \lambda} b^{\dagger2}\right) \ket{\mathrm{vac}} \, .
\end{equation}
These states have Majorana polynomials $\psi(z)$ proportional to {$_1 F_1(-J; 1/2; - \tilde{z}/4)$ when $J$ is an integer and $\tilde{z} \, _1 F_1(-J; 3/2;-\tilde{z}/4)$ when $J$ is a half-odd integer, where $\tilde{z}=- 2 \alpha^2 z^2/\lambda$, $\alpha$ is the strength of the coherent state, $\lambda =\frac{\xi}{|\xi|}\tanh|\xi|$, and $\xi$ is the strength of the squeezed state.} Since these confluent hypergeometric functions all have $J$ real roots $\tilde{z}_k<0$, all of the Majorana stars lie about the same great circle, whose azimuthal angle varies with the phases of $\alpha$ and $\xi$. We plot the Majorana constellations in a number of subspaces for such a state in Figure \ref{fig:Q}, where it is clear that the points are all distributed about the same longitudinal line. The sensitivity covariance matrix is readily calculable; for simplicity we take $\alpha^2 \xi^\ast$ to be real (and positive), making the matrix diagonal. {If the average number of excitations in each mode is large, namely $J_a=|\alpha|^2/2$ and $J_b=\sinh^2(|\xi|)/2$, we find}
\begin{equation}
    {\bm{\mathsf{C}} (\ket{\psi_{\mathrm{c+s}}} ) \simeq 
    \left (
    \begin{array}{ccc}
    \frac{J_b}{2} & 0 & 0 \\
    0 & 4 J_a J_b & 0 \\
    0 & 0 & 2J_b^2 \end{array}
    \right) \, ,}
\end{equation} 
which is useful for estimating rotations about axes in a specific plane (determined by the relative phase of $\alpha$ and $\xi$) but not axes in all three dimensions. It has been well-established that the most sensitive such states have $J_a=J_b\gg 1$ \cite{Pezze:2008aa}, for which $|\lambda|\approx 1$; this is substantiated in the $J_2$ component of $\bm{\mathsf{C}} (\ket{\psi_{\mathrm{c+s}}} )$ and reflected in the constellations for subspaces with photon numbers $\simeq |\alpha|^2$ being the most evenly distributed, per Figure~\ref{fig:Q}.
This spreading about a single great circle, which holds for all of the states in, for example, all of the states studied in Ref.~\cite{Sahota:2016aa}, shows that states from phase estimation are not immediately tailored to multiparameter estimation. }

\section{Conclusions}
{Quantum sensing has become a distinct and rapidly growing area. The case of rotations is the epitome of how quantum properties can boost sensitivity and precision.} Certain probe states are exceedingly sensitive to rotations, with the geometry of a probe state dictating its usefulness for rotation estimation. {We have made an extensive use of Majorana’s beautiful stellar representation: being purely geometric, it allows quantum mechanics to be reconciled with our physical intuition in the classical realm. This has allowed us to introduce in natural way a number of  measurement schemes} for simultaneously estimating all three parameters of a rotation and we expect these to be useful in years to come.

\ack
{We are indebted to G. Bj\"{o}rk, F. Bouchard, C. Chryssomalakos, M. Grassl, P. de la Hoz, E. Karimi, and K.  \.{Z}yczkowski for discussions.   We acknowledge financial support from the European Union Horizon 2020 (Grants ApresSF and Stormytune),  Mexican CONACYT (Grant 254127), and the Spanish MINECO (Grant PGC2018-099183-B-I00). AZG acknowledges funding from an NSERC Discovery Award Fund, an NSERC Alexander Graham Bell Scholarship, the Walter C. Sumner Foundation, the Lachlan Gilchrist Fellowship Fund, a Michael Smith Foreign Study Supplement, and Mitacs Globalink.}

\appendix
\section{Singular quantum Fisher information matrices}
\label{app:Singular QFIM}

{Singularities} are relevant to {QFI} in at least two ways. First, the density matrix may be singular, as is the case for pure states, and its rank may vary with the parameters being estimated. When this happens, the QFI suffers a discontinuity that makes it differ pointwise from the Bures metric~\cite{Safranek:2017aa,Safranek:2018aa,Sevesoetal2019}.

More physically, the QFI itself, and its matrix extension, is singular when a state cannot locally distinguish between some values of the parameters. This has been studied for {classical FI from} a Riemannian geometric perspective, where one can use the {Moore-Penrose pseudoinverse~\cite{Campbell:1991aa} to estimate a subset of the parameters with a modified {CRB}~\cite{Xavier:2004aa} (although that may constitute an overly optimistic lower bound \cite{Stoica:2001aa}). The quantum version should proceed along the same lines: {one should first identify the set of parameters that are independent, then calculate the corresponding QFI matrix for those parameters~\cite{Proctor:2018aa,Liu:2019aa}. In this Appendix we discuss} physical scenarios that lead to singular {QFI matrices} and review the implications for metrology, using rotation estimation as an illustrative example.

The QFI {matrix} depends on the set of parameters through the evolved state and the semilogarithmic derivatives. For the QFI {matrix} to be singular, there must be a basis, and therefore a parametrization, in which an entire row and column vanishes. We can analyze this occurrence in the case of pure and mixed states. 

For pure states, an entire row and column of the QFI {matrix} vanishes if and only if $\ket{\partial_\lambda\psi}\propto \ket{\psi}$ for some parameter $\lambda$. For full rank density matrices, an entire row and column {vanishes} if and only if the weights $p_i$ of each eigenstate $\ket{i}$ of the density matrix are independent of some parameter $\lambda$ and all of the eigenstates satisfy the pure states' condition $ \ket{\partial_\lambda i}\propto \ket{i}$. For sub-full rank density matrices, the latter condition is modified to only hold within the range of the density matrix: 
\begin{equation}
    \varrho  \ket{\partial_\lambda i}\propto \ket{i} \, ;
\end{equation}
either the above condition $\partial_\lambda \ket{i}\propto \ket{i}$ holds or the rank of the density matrix is changing. Since the rank of the density matrix can only change while the weights of its eigenstates are changing, a singular QFI {matrix} implies the presence of a parameter for which each eigenstate has the dependence \begin{equation}
    \ket{\partial_\lambda i}\propto \ket{i} \,.
\end{equation}

The condition on each of the eigenstates holds a) trivially, when $\ket{i}$ is explicitly independent of $\lambda$, and b) when $\ket{i}$'s global phase depends on $\lambda$. The differential equation seems to imply that the local exponential growth or decay of an eigenstate with some parameter leads to a singular QFI {matrix}, but this is precluded due to the concomitant change in the weights of the eigenstates. We will see the recurring connection between singular {QFI matrices} and global phases throughout our examples. 

A QFI {matrix} can be singular at a local set of coordinates for any state; alternatively, it can be singular for all sets of coordinates for certain states; and, finally, it may be singular for all sets of coordinates for all states. The first scenario may correspond to a coordinate singularity, the second typically heralds a shortcoming of the probe state, and the third often signifies an unseen interdependence between the parameters in question. We explore each of these scenarios in the context of rotation and related measurements.

Consider estimating the azimuthal component of a rotation axis when the polar coordinate is $0$ or $\pi$; i.e., at the north or south pole. Similarly, consider estimating the difference between the first and third Euler angles in a $ZYZ$ configuration when the second Euler angle vanishes. These points exemplify \textit{coordinate singularities}, at which one of the three rotation parameters is undefined, and so no state can be used to estimate all three because $\partial_\lambda \ket{\psi}=0$ for one of the parameters. Coordinate singularities can be identified by the QFI {matrix} being singular for all states and becoming invertible when any of the parameters is slightly perturbed or when a different parametrization with the same rank is considered. Within the original coordinate system, one can find the set of defined coordinates by the range of the QFI {matrix}; e.g., diagonalizing the {matrix} for an Euler angle parametrization shows that the null vector is associated with estimating the difference between the first and third angles \cite{Goldberg:2018aa}. Alternatively, one can find a new parametrization using a transformation whose Jacobian determinant squared cancels the singularity in the determinant. This is exemplified by switching from a spherical to a Cartesian coordinate system, for which the former's QFI {matrix} has a determinant proportional to $\theta^4\sin^2\Theta$ and the Jacobian governing the transformation has a determinant proportional to $\theta^{-2}\csc\Theta$. Linear combinations of the new parameters will also all be estimable.

Conversely, when estimating the separation between two Gaussian sources, their relative intensities, and the centroid position, the QFI {matrix} diverges as the separation vanishes \cite{Rehacek:2017ab}. This singularity is present for all states at this local point, regardless of their coherence properties \cite{Hradil:2019aa}, implying a coordinate singularity. The rub is that a Jacobian erasing the singularity requires a strange coordinate system that is not conducive to measurement: for example, a transition from the separation coordinate to the square of that coordinate. 

There are some states for which the QFI {matrix} is always singular regardless of parametrization. This is the case when estimating all three rotation parameters with classical states; {Bloch} coherent states $\ket{\bm{\mathfrak{n}}}$ cannot be used for estimating rotations about unknown axes. That {Bloch} coherent states are only sensitive to two rotation parameters is apparent from an Euler angle parametrization: by choosing the first rotation axis to coincide with the direction of the spin, the first Euler angle imparts a global phase on the state, making $\ket{\partial_\lambda\psi}=-\rmi \ket{\psi}$. One can again find this irrelevant coordinate by diagonalizing the {QFI matrix} for a classical state. It turns out that the only pure states with this issue are eigenstates of angular momentum operators $\mathbf{J}\cdot\bm{\mathfrak{n}}$, as they are the only ones for which the covariance matrix $\Cov_{\ket{\psi}}{({J}_i,{J}_j )}$ is singular, and this is the case regardless of parametrization. This immediately shows the advantage of using nonclassical states: they \textit{enable} the simultaneous estimation of all three parameters of a rotation.

Finally, some parameters were never meant to be measured. For example, in the absence of a reference phase, one cannot detect an average phase shift among all of the branches of a linear optical network. Diagonalizing the QFI {matrix} shows that only relative phases are estimable, and this holds true regardless of the probe state and of the values of the parameters being estimated; as foreshadowed, the average phase imparts a global phase to the state~\cite{Goldberg:2020ab}. Similarly, the QFI {matrix} is always singular when the parameters being estimated are not linearly independent. This scenario shows the necessity for new physics, such as the presence of an external reference phase, or new thinking, such as determining the true number of independent parameters, in order to estimate any of the parameters.

In all cases of singular QFI {matrix} we require a presciption for how to proceed with estimation. The trick detailed in Ref.~\cite{Proctor:2018aa} advises finding a parametrization in which the {matrix} is block diagonal and some maximal block is invertible, then proceeding to estimate the parameters corresponding to that block. The parameters can be inferred from a diagonalization of the QFI {matrix}, but we stress that a change in parametrization is not necessarily the same as a change of basis, as the corresponding Jacobian matrices need not be unitary. Moreover, one must avoid the use of finely tuned probe states for which preparation errors rid the QFI {matrix} of its block diagonal structure, as these cases require reintroduction of so-called nuisance parameters to the analysis. Singularities in the QFI {matrix} are essential to understanding the physical limitations of an estimation problem.

\newpage


\end{document}